\documentclass[12pt]{JHEP}
\usepackage{graphicx}

\newcommand{\beq}{\begin{equation}}
\newcommand{\eeq}{\end{equation}}
\newcommand{\beqa}{\begin{eqnarray}}
\newcommand{\eeqa}{\end{eqnarray}}

\newcommand{\tr}{{\rm Tr}}
\newcommand{\esla}{ \parbox[b]{0.6em}{$\varepsilon$} \hspace{-0.55em}
                         \parbox[b]{0.55em}{ \raisebox{-0.2ex}{$/$}}}
\newcommand{\ksla}{ \parbox[b]{0.6em}{$k$} \hspace{-0.55em}
                         \parbox[b]{0.55em}{ \raisebox{-0.2ex}{$/$}}}
\newcommand{\qsla}{\parbox[b]{0.5em}{$q$} \hspace{-0.55em}
                         \parbox[b]{0.55em}{ \raisebox{-0.3ex}{$/$}}}

\title{Large Extra Dimensions in Rare Decays
\thanks{Work supported in
part by TMR, EC-Contract No.~ERBFMRX-CT980169 (EURODAPHNE) 
and TMR,  EC-Contract No.~FMRX-CT980194 (Quantum Chromodynamics and the 
Deep Structure of Elementary Particles).
}}
\preprint{
LU TP 00-25\\
hep-ph/0006042\\
revised august 2000}
\author{Johan Bijnens and Martin Maul\\
Department of Theoretical Physics, Lund University,\\
S\"olvegatan 14A, S - 223 62 Lund, Sweden}
\abstract{
The prospect of experimental verification of
the large extra dimension scenario in rare decays is discussed.
The case of $J/\Psi$ and $\Upsilon$ to photon + missing energy
is calculated in detail, and it is shown that
the limit on the compactification 
scale $M_S$ 
lies at present 
in the ten GeV range. In contrast to the quarkonium systems,
signals in Kaon and Pion decays will be small.
}

\begin{document}
\section{Introduction}
Recently, the possibility was discussed that the compactification-scale
of large extra dimensions
could lie in the TeV range
\cite{Antoniadis:1990ew,Arkani-Hamed:1998rs,Antoniadis:1998ig,Arkani-Hamed:1999nn}, which
would offer the possibility that their effect might be visible experimentally.
For some examples using collider signatures and precision variables
see 
\cite{Cullen:2000ef,Rizzo:1999pc,Davoudiasl:2000ni,Rizzo:2000br,Antoniadis:1999bq,Graesser:2000yg}
and references therein. The generic class of models has the quarks, leptons
and other standard model fields in the usual 4 dimensions while
gravity sees the other $n$ dimensions as well. The relation between
the underlying Planck scale, $M_S$ in $n+4$ dimensions and the 4-dimensional
Planck scale, $M_P$ is then generically given in terms of the size
of the extra dimension $R$ by:
\beq
\label{planckscale}
R^n \sim M_P^2 M_S^{-(n+2)}\,.
\eeq
The papers mentioned above gave present and future experimental limits
on the scale $M_S$. In this letter we will discuss what limits can
be expected from rare decays using the missing energy signature.

As a generic choice we choose the compactified manifold to be
an $n$-dimensional torus with radius $R$. Only the $n+4$ dimensional graviton
feels this space.
The general formalism needed
is discussed in \cite{Han}. The lowest order coupling of the 4-dimensional
states from the $n+4$-dimensional graviton,
the 4-dimensional graviton, $\tilde{h}_{\mu\mu^\prime}$,
the dilaton, $\tilde\phi$,  and their Kaluza-Klein excitations (KK),
is model-independent and can be written in terms of the energy-momentum
tensor.
This approximation is non-renormalizable and the resulting expressions
can only be used as an effective
theory. Some loop-diagram calculations have been performed as well,
e.g. \cite{Graesser:2000yg,Akhoury:2000pk}. 

Rare decay measurements take place at much lower momenta than the
collider limits discussed earlier. The main question is whether
the larger precision obtainable in these experiments can compensate
for the lower scales involved. Let us first look at a generic decay
of a particle with mass $M_D$ and check
what we could expect on dimensional grounds.
Phase space factors are of similar magnitude so we neglect these
in this argument.
The couplings of gravitons and KK-excitations are all proportional
to $\kappa=\sqrt{16\pi G_N}$, a decay-width is thus generically suppressed
by a factor of $\kappa^2 M_D^2$ compared to the usual decays.
For scales $R\gg1/M_D$ there is an additional enhancement factor
due to the total number of KK-excitations that exists. For the simple
compactification discussed above this number is \cite{Han}:
\beq
\label{numberKK}
N_{m_{\tilde h}\leq M_D}
= \int_0^{M_D^2}dm^2 \frac{R^n m^{n-2}}{(4\pi)^{n/2}\Gamma(n/2)}\,.
\eeq
Therefore a generic branching fraction is of order:
\beq
\label{generic}  
R^n \kappa^2 M_D^{n+2}\approx \left(\frac{M_D}{M_S} \right)^{n+2}
\;.
\eeq
Notice that this argument neglects all other dimensionful factors
except phase-space going into the other decays, these can of course only
be included for specific decays and we discuss the case of quarkonium in
detail below.  Eq.~(\ref{generic}) indicates that the sensitivity to 
large extra dimensions is stronger when heavy mesons are probed, thus
motivating to study heavy quarkonia like $J/\Psi$ and $\Upsilon$.
In addition their main decay is 3-body and proportional to
$\alpha_S^3$ while the signal is two-body and proportional to $\alpha$,
providing an extra enhancement factor for the quarkonia decays.
Eq.~(\ref{generic})
shows also that the sensitivity decreases the more extra dimensions
are 'active'.

The impact of large extra dimensions discussed here follows the
formalism of \cite{Han} throughout. The possible existence
of universal torsion-induced interaction from large extra dimensions
\cite{Chang:2000yw}  might lead to four-quark vertices which could
enhance the sensitivity of rare meson decays 
to large extra dimensions significantly.

In the next section we present some arguments mainly based on angular
momentum and helicity conservation why we expect,
in contrast to heavy quarkonia,  Kaon and Pion decays to be less
promising candidates. Some of these arguments are also
applicable to $B$ and $D$-decays.
In Sect. \ref{JPSI} we discuss quarkonium decays
into photon + KK-excitation in detail. We present explicit results for
$J/\Psi$ and $\Upsilon$, including the total branching ratios
and the photon energy spectrum.

\section{Kaon and Pion Decays}

\subsection{Angular Momentum and Helicity}

A generic problem with decays to gravitons is that they are
spin two. As a consequence for most decays there needs to be
a component of angular momentum in the final state as well.
This leads to typical additional suppression coming from the matrix
elements. This argument is not quite so strong for the dilaton which
has spin-0, but since a lot of the options require the other particle to
be a photon similar arguments apply.

This is the case for example in $\pi^0$ decays where calculations
of $\pi^0\to\gamma\tilde\phi$ and $\pi^0\to\gamma\tilde h$ from
triangle diagrams in quark models all give zero to leading order.

\subsection{$K$ to $\pi$ KK-excitation}

Weak $K$ decays are in first approximation well described by
chiral Langrangians expressed in terms of point-like meson fields
with the generic structure\footnote{Higher order terms can be brought
in a similar form near the mass-shell by using the equations of motion}:
\beq
\label{chirallag}
\sum_{ij}\left(a_{ij}\partial_\mu M_i\partial^{\mu}M_j-m^2_{ij}M_i M_j 
+{\cal O}(M_i^3)
\right)\,.
\eeq
Here $M_i$ is a generic particle field.
The KK-excitation couplings are then obtained by calculating the
energy momentum tensor from Eq.~(\ref{chirallag}).

As we can always diagonalize the quadratic terms in Eq.~(\ref{chirallag}),
no $K\to\pi$ transition exists and thus the leading
order contribution
to $K\to\pi\tilde\phi$ or $K\to\pi\tilde h$ vanishes as well.
Decays with more particles in the final state are of course permitted.
A possible nonvanishing source comes from Penguin-like diagrams
with external gravitons attached. We can  expect the
extra loop factors involved in this case to provide extra suppressions.

\subsection{A naive argument for $B$ and $D$ decays}

The above argument of minimally coupled KK-excitations is also
present when we consider them as coming from underlying $B\to P$,
with
$P$ some particle, transitions.
On the other hand, since momenta in the final state here are much larger
this might not lead to quite so strong suppressions as in the Kaon-Pion case.

Contributions from gravitonic penguins are however subject to the same
suppressions as normal Penguins so we can expect them to be
at most comparable to the $b\to s\gamma$ transition as a starting point. 

\section{Quarkonium}
\label{JPSI}

We now calculate the decays of quarkonium to a photon and a
KK-excitation. We restrict ourselves to the lowest $S$-wave states
and treat the quarks in the static approximation. An overview of this
type of calculations can be found in \cite{Buchmuller}. The precise
formalism of helicity projections that we use is that of \cite{Kuhn}.
In this approximation the quarks are considered to be at rest and thus
no extra angular momentum can be produced. The angular momentum and
helicity arguments of the previous section thus apply as well and
we find consequently that the spin zero states, $\eta_c$ and $\eta_b$
do not decay to photon--KK-excitation to leading order.

We have chosen decays including a photon since they provide a clean signature
and are unambiguously calculable. Hadronic decays require at least two-gluons
in addition to the graviton so they will be of relative order
$\alpha_S^2/\alpha$ with extra color and phase-space factors.
We do not expect them to be of a very different order of magnitude than
the ones we calculated.

\subsection{The Calculation}

\FIGURE{
\includegraphics[width=12cm]{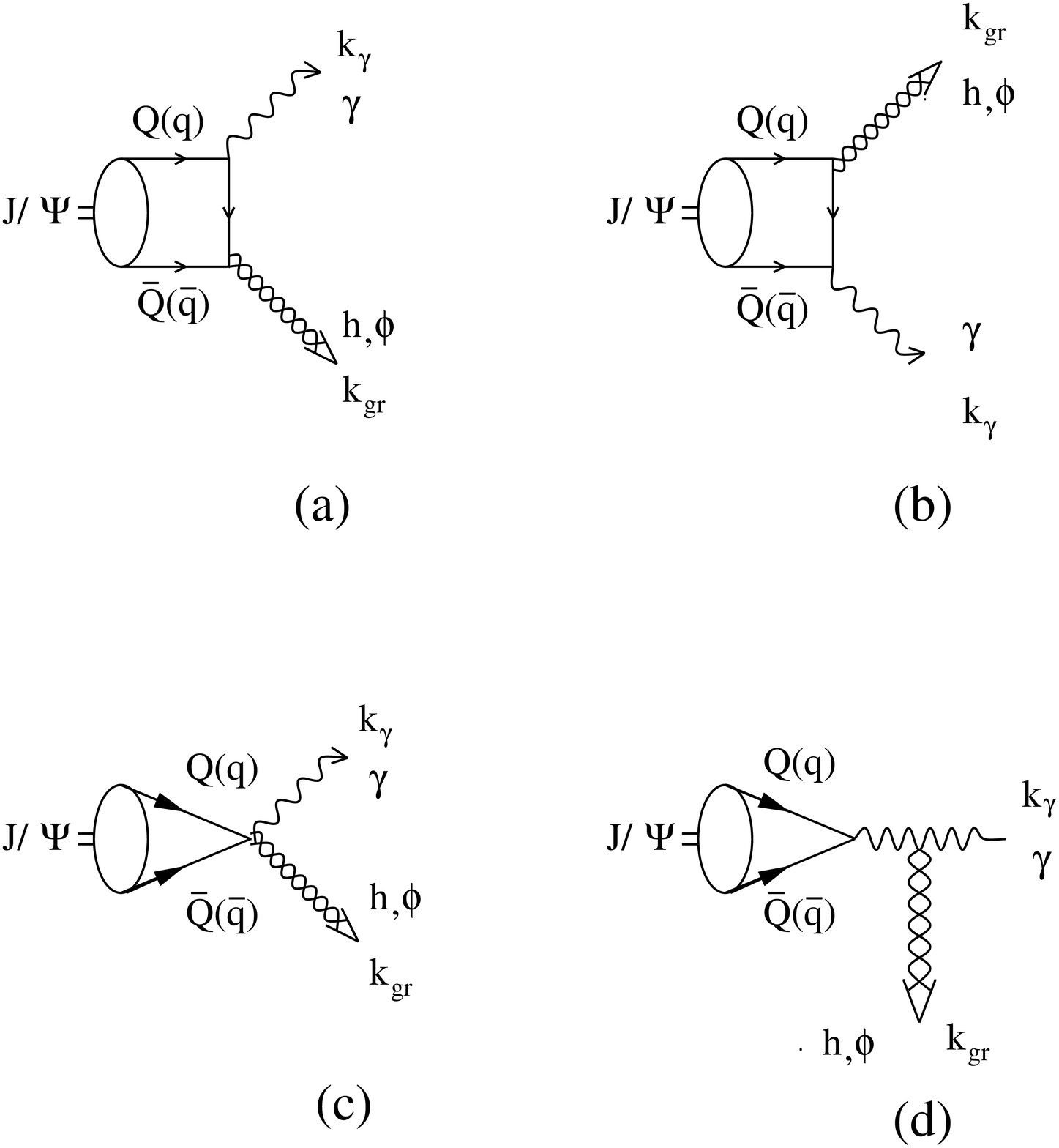}
\caption{Diagrams contributing to quarkonium decays into a photon and 
a KK-excitation.}
\label{fig1}}

The diagrams
 contributing to the quarkonium decay into a photon and a KK-excitation
are given in Fig.~\ref{fig1}. For definiteness, we concentrate here
in the calculation on $J/\Psi$ decay. The replacements necessary for
the calculation of the $\Upsilon$ decay are obvious.

The kinematics is the annihilation reaction
of the quarkonium state, $\mbox{Q}\bar{\mbox{Q}}$
into a photon, $\gamma$ and a graviton $\tilde h$ or dilaton $\tilde\phi$.
Basically, this is the reaction
${\mbox{Q}}( q) + \bar{\mbox{Q}}(\bar q)
\to \gamma(k_\gamma) + (\tilde h,\tilde\phi)(k_{\mbox{\tiny gr}})$ for 
quarks at rest projected on the quarkonium state.
Using the projection on charmonium states, as
described in \cite{Kuhn}, and the Feynman rules
for the coupling of the KK-gravitons with massive fermions
\cite{Han}, we get for the spin-2 KK contribution:
\beqa
\Psi^{L_z S_z}_{(a)\mu\mu'\nu}({\bf q}) &=& 
-i\frac{ N_c e e_q\kappa}{2\sqrt{2}(2 m_c)}  \Psi_{L L_z}({\bf q})
\bar q_{\mu'}
\frac{\tr\left[ (\qsla+m_c) \esla(S_z)(\bar {\qsla} -m_c)
\gamma_\mu(\ksla_{\rm gr}-\bar {\qsla} + m_c)\gamma_\nu\right]}
{(k_{\rm gr} - \bar {q})^2 - m_c^2}
\nonumber \\
\Psi^{L_z S_z}_{(b)\mu\mu'\nu}({\bf q}) &=& 
i\frac{N_c e e_q\kappa}{2\sqrt{2}(2 m_c)}  \Psi_{L L_z}({\bf q})
 q_{\mu'}
\frac{\tr\left[ (\qsla+m_c) \esla(S_z)(\bar {\qsla} -m_c)
\gamma_\nu( {\qsla}-\ksla_{\rm gr}+ m_c)\gamma_\mu\right]}
{(k_{\rm gr} - {q})^2 - m_c^2}
\nonumber\\
\Psi^{L_z S_z}_{(c)\mu\mu'\nu}({\bf q}) &=& 
-i\frac{N_c e e_q\kappa}{2\sqrt{2}(2 m_c)}  \Psi_{L L_z}({\bf q})
\tr\left[ (\qsla+m_c) \esla(S_z)(\bar {\qsla} -m_c)
\gamma^\rho\right]\eta_{\mu\nu}\eta_{\mu'\rho}
\nonumber\\
\Psi^{L_z S_z}_{(d)\mu\mu'\nu}({\bf q}) &=& 
2i\frac{N_c e e_q\kappa}{2\sqrt{2}(2 m_c)}  \Psi_{L L_z}({\bf q})
\frac{\tr\left[ (\qsla+m_c) \esla(S_z)(\bar {\qsla} -m_c)
\gamma^\rho\right]}
{(2m_c)^2}\nonumber \\
&& \times \left[
Q \cdot k_{\gamma}\eta_{\mu\rho}\eta_{\mu'\nu} 
-\eta_{\mu\nu} Q_{\mu'}(k_{ \gamma})_\rho
-\eta_{\mu\rho}Q_\nu(k_{\gamma})_{\mu'}
+\eta_{\rho\nu} Q_\mu(k_{\gamma})_{\mu'}\right]\;.
\eeqa 
For definiteness we have written the quark mass as $m_c$ and used
$2 m_c = M_\Psi$, with $\Psi$ the quarkonium state.
The superscript indicates the orbital and spin $z$-component and the
subscripts are 
the indices coupling to the KK-graviton, $\mu\mu^\prime$ and
to the photon, $\nu$. 
$Q = q+\bar q$ and $\Psi_{LL_z}({\bf q})$ is the quarkonium
wave function with orbital angular momentum $L$ and its z-component
$L_z$. The quark and antiquark momentum are on-shell with
spatial component {\bf q} and $-${\bf q} respectively.
$\varepsilon(S_z)$ is the polarization amplitude for the quarkonium
state. For $\eta_c$ and $\eta_b$, $\esla(S_z)$ should be replaced by
$\gamma_5$.
For the sum of all four amplitudes we write:
\beq
 \Psi^{L_z S_z}_{\mu\mu'\nu}({\bf q}) = 
 \Psi^{L_z S_z}_{(a)\mu\mu'\nu}({\bf q})
+\Psi^{L_z S_z}_{(b)\mu\mu'\nu}({\bf q})
+\Psi^{L_z S_z}_{(c)\mu\mu'\nu}({\bf q})
+\Psi^{L_z S_z}_{(d)\mu\mu'\nu}({\bf q})\;.
\eeq 
In the static limit 
${\bf q }\to 0$ the amplitudes simplify further
and we get,
summing over the photon and graviton polarizations,  for the 
squared amplitude of the $J/\Psi$ ($^1S_3$ state):
\beqa
|M_{\tilde h}^{(J/\Psi)}|^2 &=& \int\frac{d^3\bf q}{(2\pi)^3}
          \int\frac{d^3\bf q'}{(2\pi)^3}
      \Psi^{Lz=0 S_z}_{\mu\mu'\nu}({\bf q})
\left(\Psi^{L_z=0 S_z}_{\mu_1{\mu_1}'\nu'}({\bf q'})\right)^*
 B^{\mu\mu',\mu_1{\mu_1}'} \left(-\eta^{\nu\nu'} \right)
\nonumber \\
&=& \frac{N_c}{4}\alpha_{\rm em} \kappa^2 R_0^2 e_q^2 m_c
\left\{ \begin{array}{ll}
          4\frac{m_{\bf n}^2}{m_c^2}\;; & {\rm if}\; S_z = 0 \\
           \frac{8}{3} + \frac{m_{\bf n}^4}{m_c^4}\;; & {\rm if}\; S_z = \pm 1
         \end{array} \right.\,.
\eeqa
We have chosen the spin quantization axis parallel to the photon momentum.
The polarization tensor of the spin-2 KK states is given by
\cite{Han}:
\beq
B_{\mu\nu,\rho\sigma}^{\tilde h}
=        \tilde\eta_{\mu\rho}\tilde \eta_{\nu\sigma}
            +\tilde\eta_{\mu\sigma}\tilde \eta_{\nu\rho }
-\frac{2}{3} \tilde\eta_{\mu\nu}\tilde \eta_{\rho\sigma}
\;,\quad
\tilde \eta_{\mu\nu} = 
\left(\eta_{\mu\nu} -\frac{(k_{\rm gr})_\mu (k_{\rm gr})_\nu}
{m^2_{\bf n}}\right)\; .
\eeq 
The normalization of the $J/\Psi$ wave function is given by:
\beq
\int\frac{d^3{\bf q}}{(2\pi)^3} \Psi_{00}({\bf q}) = 
\frac{R_0}{\sqrt{4\pi m_c N_c}}\;.
\eeq
The constant $R_0$ describes the $S$-wave function at the origin.
For the spin-0 KK state (dilaton) the amplitudes read:
\beqa
\Psi^{L_z S_z}_{(a)\nu \; ij}({\bf q}) 
&=& -i\delta_{ij}\frac{N_c e e_q\kappa\omega}
{\sqrt{2}(2 m_c)}  \Psi_{L L_z}({\bf q})
\nonumber\\&&\times
\frac{
\tr\left[ (\qsla+m_c) \esla(S_z)(\bar {\qsla} -m_c)
\left(-\frac{3}{2}\bar{\qsla}+\frac{3}{4}\ksla_{\rm gr}-2m_c\right)
(\ksla_{\rm gr} -\bar{\qsla} + m_c)
\gamma_\nu\right]
}
{(k_{\rm gr} - \bar q)^2 - m_c^2}
\nonumber \\
\Psi^{L_z S_z}_{(b)\nu \; ij}({\bf q}) 
&=& -i\delta_{ij}\frac{N_c e e_q\kappa\omega}
{\sqrt{2}(2 m_c)}  \Psi_{L L_z}({\bf q})
\nonumber\\&&\times
\frac{\tr\left[ (\qsla+m_c) \esla(S_z)(\bar {\qsla} -m_c)
\gamma_\nu
(\qsla - \ksla_{\rm gr} + m_c)
\left(\frac{3}{2}{\qsla}-\frac{3}{4}\ksla_{\rm gr}-2m_c\right)
\right]}
{(k_{\rm gr} - q)^2 - m_c^2}
\nonumber \\
\Psi^{L_z S_z}_{(c)\nu \; ij}({\bf q}) &=& 
i\delta_{ij}\frac{3}{2}\frac{N_c e e_q\kappa\omega}{\sqrt{2}(2 m_c)}
  \Psi_{L L_z}({\bf q})
\tr\left[ (\qsla+m_c) \esla(S_z)(\bar {\qsla} -m_c) \gamma_\nu\right]
\nonumber \\
\Psi^{L_z S_z}_{(d)\nu \; ij}({\bf q}) &=& 0\;.
\eeqa 
For the sum of all three nonzero amplitudes we write:
\beqa
 \Psi^{L_z S_z}_{\nu \; ij}({\bf q}) &=& 
 \Psi^{L_z S_z}_{(a)\nu \; ij}({\bf q})
+\Psi^{L_z S_z}_{(b)\nu \; ij}({\bf q})
+\Psi^{L_z S_z}_{(c)\nu \; ij}({\bf q})\;.
\eeqa 

Then the amplitude squared for the spin-0 KK state (dilaton) 
for the  the $J/\Psi$ ($^1S_3$ state) is given by:
\beqa
|M_{\tilde \phi}|^2 &=& \int\frac{d^3\bf q}{(2\pi)^3}
          \int\frac{d^3\bf q'}{(2\pi)^3} 
      \Psi^{L_z=0 S_z}_{\nu \; ij}({\bf q})
B_{ij,i'j'}^{\tilde\phi}
\left(\Psi^{L_z=0 S_z}_{\nu' \; i'j'}({\bf q'})\right)^*
\left(- \eta^{\nu\nu'} \right) 
\nonumber \\
&=& 2(n-1)N_c\alpha_{\rm em} \kappa^2 \omega^2 R_0^2 e_q^2 m_c\;,
\eeqa
for $S_z=\pm1$ and zero for $S_z=0$. The expression $B_{ij,i'j'}^{\tilde\phi}$
comes from the dilaton propagator and has the form:
\beq
 B_{ij,i'j'}^{\tilde\phi} = 
\frac{1}{2}\left( P^{\vec n}_{ii'}P^{\vec n}_{jj'} 
               +  P^{\vec n}_{ij'}P^{\vec n}_{ji'}\right)
\;.
\eeq
And the projectors $P^{\vec n}_{ii'}$ are defined through:
\beq
 P^{\vec n}_{ii'} = \delta_{ii'}-\frac{n_i n_{i'}}{\vec n^2}\;;
\quad  
 P^{\vec n}_{ij} P^{\vec n}_{jk} = P^{\vec n}_{ik}\;;
\quad
 P^{\vec n}_{ii} = n-1\;.
\eeq 
A simple check is that all amplitudes given above vanish for the $\eta_c$.
The decay rate for the (unpolarized) $J/\Psi$ is then given averaging
over the initial possible polarization and the summation over the
KK modes of masses $m_{\bf n}$
as described in \cite{Han} and Eq.~(\ref{numberKK}):
\beqa
\Gamma(J/\Psi\to \gamma+gr)_{\tilde h, \tilde \phi} &=& \frac{1}{3}\sum_{S_z}
\frac{1}{2M_{J/\Psi}}\sum_{\bf n} 
\int\frac{d^4k_\gamma}{(2\pi)^3}
\int\frac{d^4k_{gr}}{(2\pi)^3} 
\delta^+(k_\gamma^2)
\delta^+(k_{gr}^2 -m_{gr}^2)
\nonumber \\ && \times 
(2\pi)^4 \delta(k_{J/\Psi}-( k_\gamma + k_{gr}))
 |M(m_{\bf n}^2,S_z)_{\tilde h, \tilde \phi}|^2
\nonumber \\
&=& \frac{1}{3}\sum_{S_z}
 \frac{1}{32\pi m_c^2 } \int_0^{(2m_c)^2}
 \frac{dm^2_{\bf n}R^n m_{\bf n}^{n-2}}{(4\pi)^{n/2}\Gamma(n/2)}
\left(
\frac{(2m_c)^2-m_{\bf n}^2}{4m_c}
\right) 
\nonumber \\ && \times
|M(m_{\bf n}^2,S_z)_{\tilde h, \tilde \phi}|^2 \;.
\eeqa
The $+$ sign over the delta functions means that the energy must be positive. 
The decay width becomes finally:
\beqa
\Gamma(J/\Psi\to \gamma+gr)_{\tilde \phi} &=& (n-1)
\frac{\alpha_{em}\kappa^2\omega^2 R_0^2 N_c e_q^2}{24\pi m_c } \int_0^{(2m_c)^2}
 \frac{dm^2_{\bf n}R^n m_{\bf n}^{n-2}}{(4\pi)^{n/2}\Gamma(n/2)}
\left(
\frac{(2m_c)^2-m_{\bf n}^2}{4m_c}
\right)
\nonumber \\
\Gamma(J/\Psi\to \gamma+gr)_{\tilde h} &=& 
\frac{\alpha_{em}\kappa^2 R_0^2 N_c e_q^2}{192\pi m_c } \int_0^{(2m_c)^2}
 \frac{dm^2_{\bf n}R^n m_{\bf n}^{n-2}}{(4\pi)^{n/2}\Gamma(n/2)}
\left(
\frac{(2m_c)^2-m_{\bf n}^2}{4m_c}
\right) 
\nonumber \\ && \times
\left[\frac{8}{3} + 2 \left(\frac{m_{\bf n}}{m_c}\right)^2 
+ \left(\frac{m_{\bf n}}{m_c}\right)^4\right] 
\;.
\eeqa
The photon spectrum can be easily deduced from the formulas given above
using the kinematical relation
$m_{\bf n}^2 = M_{J/\Psi}^2 - 2 M_{J/\Psi} E_\gamma
= (2m_c)^2-4 m_c E_\gamma$.
The integrals are all polynomials and can be done explicitly.

\subsection{Numerical Results}

\FIGURE{
\includegraphics[width=15cm]{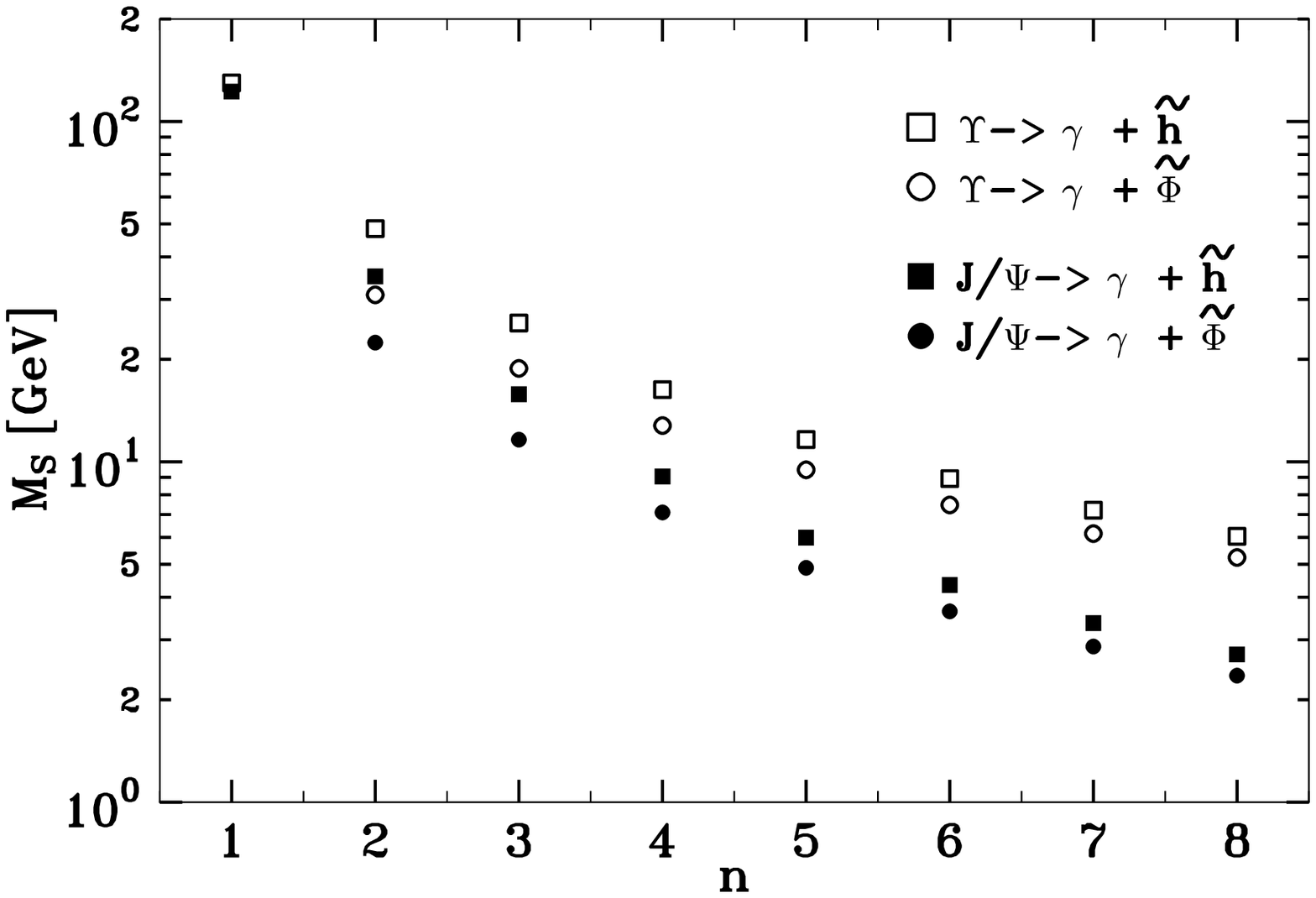}
\caption{The lower limits on the  scale $M_S$ from the decays
 $J/\Psi\to\gamma\tilde h$, $J/\Psi\to\gamma\tilde\phi$,
 $\Upsilon\to\gamma\tilde h$ and $\Upsilon\to\gamma\tilde\phi$
as a function of the number of extra dimensions $n$.}
\label{figlimits}
}

For the numerical results we set $\alpha_{\rm em} = 1/137$,
$e_q = 2/3$, 
$\omega = \sqrt{\frac{2}{3(n+2)}}$, 
$2 m_c = M_{J/\Psi} = 3097$ MeV, 
$\kappa = \sqrt{16\pi G_N}$, and
the gravitation constant $G_N = 6.70711\times 10^{-45} \;{\rm MeV}^{-2}$.
The value for $R_0$ can be obtained in the accuracy of our calculation
from the tree level width for the decay $J/\Psi\to e^+e^-$ 
\cite{PDG}:
\beq
\Gamma(J/\Psi \to e^+e^-) = 
\frac{4 \alpha_{\rm em}^2 e_q^2 N_c}{3 M^2_{J/\Psi}} R_0^2 = 5.2374 \;{\rm keV}
\;.
\eeq
This relation receives sizable QCD  corrections which is the
reason for the fact that  the
value of $R_0$ is not precisely known, see e.g. the discussion
in \cite{Eichten:1995ch}.
Similar corrections can
be expected for the decays into photons and KK-excitations.
These corrections will affect the precise values of the 
numbers discussed below, but not any of the 
conclusions.

\FIGURE{
\includegraphics[width=15cm]{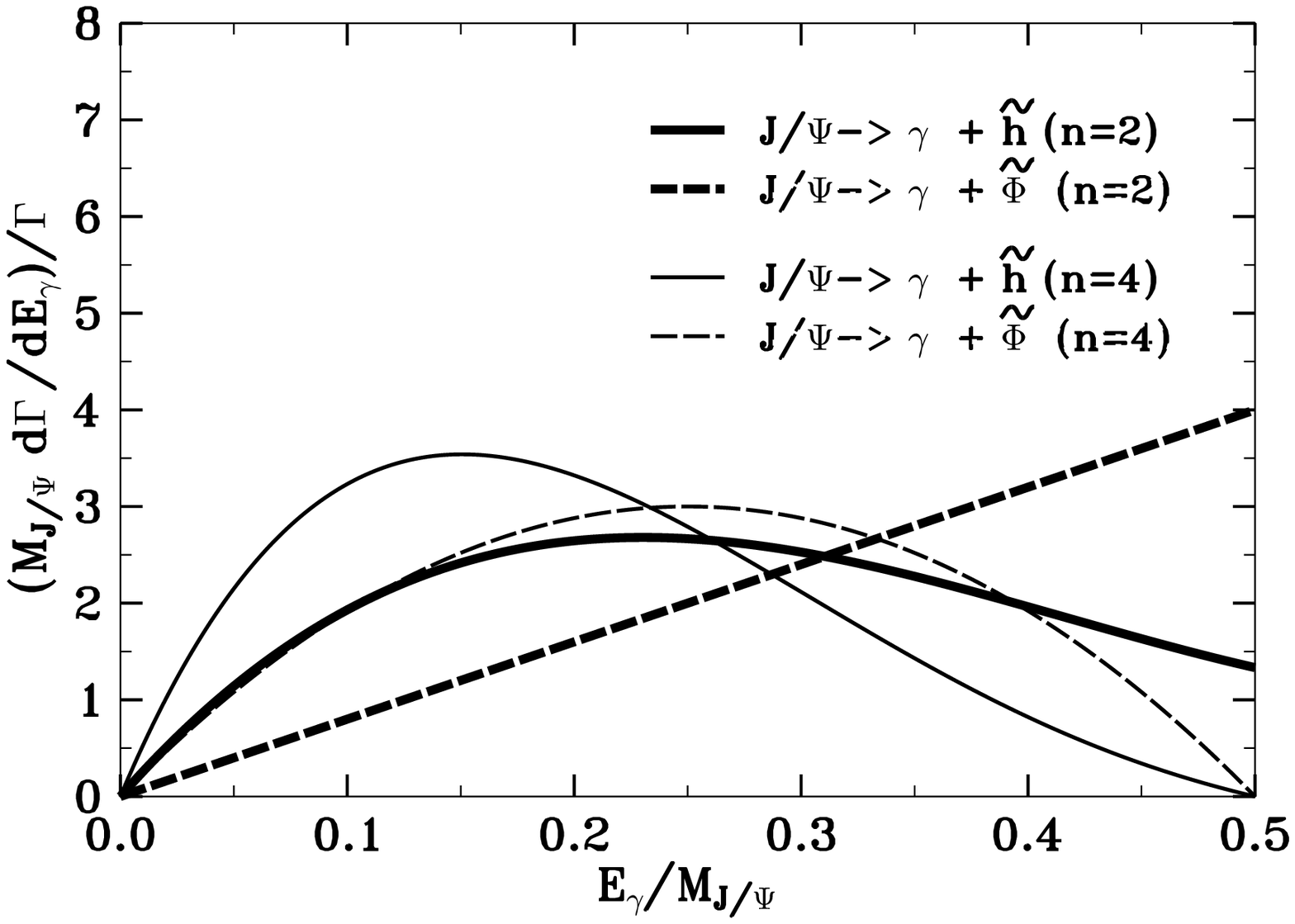}
\caption{The photon spectrum for the dilaton $\tilde\phi$
and graviton $\tilde h$ case.
Shown are $n=2,4$.
\label{figspectrum}}
}

We can now define a critical compactification scale, when the contribution
from graviton radiation becomes as big 
as the precision reached in today's experiments looking at
rare $J/\Psi$ decays.
{}From the data given in \cite{PDG} we expect that 
a Branching ratio of:
\beq
B(J/\Psi\to \gamma + \tilde h, \tilde\phi, R_{\rm crit})
= 10^{-5} 
\eeq
would be easily measurable.
We used $\Gamma_{J/\Psi} = 87$~keV to normalize
the branching ratios.
The resulting  $R_{\rm crit}$, 
converted into a limit on $M_S$ using Eq.~(\ref{planckscale})
with $M_P = \sqrt{1/G_N}$ ,
is shown in Fig. \ref{figlimits} as a function of $n$.
The limits from $J/\Psi\to\gamma\tilde h$ and $J/\Psi\to\gamma\tilde\phi$
are shown as well as the similar limits from $\Upsilon\to\gamma\tilde h$
and $\Upsilon\to\gamma\tilde\phi$.
For the latter ones
 we used $M_\Upsilon=9.460$~GeV, $\Gamma(\Upsilon \to e^+e^-) = 1.323$~keV,
$\Gamma_\Upsilon = 52.5$~keV \cite{PDG}, $e_q = -1/3$ and a measurable
branching ratio of $10^{-4}$. It is seen that in general the limit
on $M_S$ is larger for the $\Upsilon$ decay than for the $J/\Psi$ decay and
that the spin-2 KK state dominates over the spin-0 KK state  (dilaton).
In general the critical values for $M_S$ lie in the ten GeV range.

The photon spectrum as a function of $E_\gamma/M_{J/\Psi}$
is shown in Fig. \ref{figspectrum} for the spin-2, $\tilde h$,  and spin-0
case $\tilde\phi$. It shows a characteristic shape for 
$\tilde h$ and $\tilde\phi$
depending on $n$.
Notice that the relative spectrum shown
is neither dependent on $R$ nor on $M_{J/\Psi}$.

\section{Summary and Conclusions}
\TABLE{
\begin{tabular}{|c||r|r|r|}
\hline &&& \\
 Experiment & $M_S(n=2)$ & $M_S(n=4)$ & $M_S(n=6)$ \\
&&&\\
\hline\hline
&&&\\
DELPHI & 1250.0  & 790.0 & 590.0 \\
$\Upsilon \to \gamma+\tilde h$ & 48.4 & 16.3 & 8.9 \\
&&& \\
\hline
\end{tabular}
\caption{ Limits on $M_S$ in GeV from missing-energy processes.
The numbers for DELPHI are taken from \cite{LandsbergTalk}.}
\label{comparision}
}

We have presented general arguments why looking for large extra dimensions
in rare meson decays in  cases where the 
missing energy is carried away by KK-excitations
of the graviton in $4+n$-dimensions is most promising for heavy
quarkonia systems.
These included general dimensional arguments, angular momentum conservation
and arguments due to the pointlike structure at relevant scales.

As an example,
we computed $J/\Psi$ and $\Upsilon$ decays into a photon and KK-excitations. 
The  limits to be expected from these processes 
lie in the  GeV range. The spectrum of the photon-energy $E_\gamma$
in those decays is characteristic for the number $n$ of extra large dimensions
and different for spin-2 and spin-0 KK states. In general, 
the spin-2 KK graviton yields larger contributions than the spin-0 dilaton.
$\Upsilon$ decays are
more favorable than $J/\Psi$-decays due to the larger mass involved
and the possibility of improving the branching limit considerably at
the $B$-factories.

One can compare the limits on $M_S$ in rare decays with the
present best limits from $e^+e^-$ colliders given by the DELPHI
collaboration,  see Tab.~\ref{comparision}. (Limits from
other present experiments, see \cite{LandsbergTalk} for details, 
are of the same order.) The values from DELPHI 
are  better by one or two orders of magnitude than
what can be obtained from rare decays with present limits.
However,  our rare decays test mainly the couplings to heavy quarks 
while the high-energy
colliders test mainly couplings to gauge bosons and light fermions,
so qualitatively
there is a difference between what is tested in rare decays and
what is tested by lepton or even hadron colliders like the LHC.
Moreover one should take into account that through possible
improvements at B-Factories the limits on large extra dimensions
could be possibly increased by an order of magnitude. 

\section*{Acknowledgments}
We wish to thank P.~Di Vecchia for fruitful discussions. This work was
supported in part by TMR, EC-Contract No.~ERBFMRX-CT980169 (EURODAPHNE) 
and TMR,  EC-Contract No.~FMRX-CT980194 (Quantum Chromodynamics and the 
Deep Structure of Elementary Particles).

\end{document}